# Extending Term Suggestion with Author Names

Philipp Schaer, Philipp Mayr, Thomas Lüke

GESIS – Leibniz Institute for the Social Sciences,
Unter Sachsenhausen 6-8, 50667 Cologne, Germany
`{philipp.schaer|philipp.mayr|thomas.lueke}@gesis.org`

**Abstract.** Term suggestion or recommendation modules can help users to formulate their queries by mapping their personal vocabularies onto the specialized vocabulary of a digital library. While we examined actual user queries of the social sciences digital library Sowiport we could see that nearly one third of the users were explicitly looking for author names rather than terms. Common term recommenders neglect this fact. By picking up the idea of polyrepresentation we could show that in a standardized IR evaluation setting we can significantly increase the retrieval performances by adding topical-related author names to the query. This positive effect only appears when the query is additionally expanded with thesaurus terms. By just adding the author names to a query we often observe a query drift which results in worse results.

## 1 Introduction

When we look at specialized information systems like scientific digital libraries a long-known retrieval-immanent problem becomes clear: The same information need can be expressed in a variety of ways. This is especially true for scientific literature. Each scientific discipline has its own domain-specific language and vocabulary. For a long time indexers and digital library curators tried to encode this language into documentary tools like thesauri or classification systems, which are used to describe scientific documents. When we think of information retrieval as "fundamentally a linguistic process" like Blair [1] did, users have to be aware of these specialized documentation systems and their vocabularies. In order to get the most relevant results users should formulate their queries to best match these controlled thesauri terms. So-called search term recommenders (STR) try to map any search terms to controlled vocabulary terms [2] to support the user in the query formulation phase.

But terms from the title, abstract, full texts or controlled keywords are not the only entities that represent a digital library's document. Typical digital library systems incorporate many additional metadata representations of documents like author names, affiliation etc. These entities can be used during search but mostly in the form of free text search or during an "extended search". This pattern can be seen when studying the query formulations of actual users. We conducted a log file analysis with the social sciences digital library Sowiport[1]. Here we can see that in the 1000 most popular user queries nearly 1/3 of the users are explicitly searching for persons and only 1/5 are using controlled terms. Despite the fact that more users are looking for people rather than controlled terms, the included search term recommender in Sowiport (and the ones implemented in other DL systems) neglects this.

Table 1. Query log analysis for the social science portal Sowiport. We analyzed the 1000 most popular user queries (n=129,251). Last line ("others") combines year, institution and location.

| Search entities | Number of queries | % of total queries |
| --- | --- | --- |
| Any Entity (all fields) | 58,754 | 45,5% |
| Person(s) | 40,979 | 31,7% |
| Keywords | 26,959 | 20,9% |
| Title | 1,108 | 0,9% |
| Source | 581 | 0,4% |
| Others | 354 | 0,3% |

Therefore in this paper we argue for the explicit use of polyrepresentation of documents and different inter-document features to give a more complete term suggestion during query formulation. We present a general approach to obtain feasible author suggestions using simple co-occurrence analysis methods.

The paper is structured as follows: We give an overview on recent works in the field and the application of search term recommendation, expert finding and polyrepresentation in section 2. Section 3 presents an outline on how to make use of formerly unused metadata fields like author information in form of search term recommenders. This approach will be evaluated in section 4 using standard IR evaluation methods. We will conclude in section 5 with a general argumentation on why to develop the idea of polyrepresentation-aware recommenders further.

## 2 State of the Art in Digital Library Recommenders

In the following section we will give a very short overview on the field of term and author recommending systems in the domain of digital libraries and on the principle of polyrepresentation and its application.

### 2.1 Recommender in Digital Libraries

Recommending useful entities (e.g. terms) for search or browsing is a standard feature in typical search systems, and also in scientific DL. A couple of methodological different recommenders have been published in the DL domain. In the following we outline three typical examples extending the search process with additional document features or value-added services.

---

[1] http://www.gesis.org/sowiport/en/

Geyer-Schulz et al. [3] presented a recommender system based on log file analysis, which they implemented in a legacy library system. Their approach was based on the repeat-buying theory and could present related documents to a given document ("others also use…"), newer approaches extended their work [4]. Hienert et al. [5] presented another recommending approach, which focused on the mapping problem between specialized languages of discourse and documentary languages. General query terms and phrases were mapped onto a controlled vocabulary. Users could choose from these term suggestions during the query formulation phase to extend the initial user query. A third approach is called expert recommendation or expert search [6, 7]. The general idea is to support scientists in finding relevant literature from related peers. These recommended scholars don't necessarily have to be known to a searcher since it's a desired outcome to find new people's work.

Most of these systems allow the user to select some of their proposed entities (related documents, terms or author names) and most authors claim that their approaches help user in the search process, open alternative search paths and would optimize search systems performance. Nevertheless with few exceptions [8] these systems are not included in live retrieval systems (e.g. as query expansion mechanisms).

## 2.2 Polyrepresentation

The principle of polyrepresentation is described by Ingwersen and Järvelin [9]: "[…] the more interpretations of different cognitive and functional nature […] that point to a set of objects in so-called cognitive overlaps, and the more intensely they do so, the higher the probability that such objects are relevant (pertinent, useful) to the information (need) situation at hand […]".Cognitive differences in the representation derive from the interpretations from different actors, like keywords given by the author contrary to professional indexers keywords. A functional difference can – but does not have to – derive from the same actor and incorporate "contextual properties" like journal name, publication year, country, and others. Contrary to the principle of redundancy the principle of polyrepresentation focuses on the usage of cognitive and functional differences rather than to avoid them. It aims at a broad leverage in effectiveness by using and combining these differences compared to the single usage of each of the representations. Furthermore the idea is to surpass the drawbacks and weaknesses of some representations.

The general feasibility of this approach was shown in different studies like the one from Skov et. al [10]. They did query experiments using the Cystic Fibrosis data set extracted from Medline. In their work they transferred an initial query into different query representations/fields like title, abstract, MeSH major and minor controlled terms. While using a rather small document set with only 1,239 documents and 29 topics they could nevertheless show a significant improvement in retrieval effectiveness measured in precision, recall and cumulative gain. The approach from Skov is common in the domain of query term suggestion and query expansion: Query terms are expanded with semantically close terms from a thesaurus or other vocabulary. So a query on *What are the hepatic complications of cystic fibrosis?* is expanded with terms like *liver* or *hepatectomy*.

## 3      An Author-aware Term Suggestion and Expansion Module

According to Larsen [11] the "polyrepresentation of the seeker's cognitive space is to be achieved by extracting a number of different representations from the seeker". He lists three representations or intentions of a seeker: a "what", a "why" and a work task or domain description. We argue that there are many other representations of a seeker's intention or information need. One of these might be a "who", meaning the potential desire to get results from a specific person or a person with reputation, knowledge or presence in the field. While a lot of work has been put into the exploitation of the functional polyrepresentation we argue that this cognitive site of a seeker's intention is ignored too often.

The general idea presented in this paper is to exploit the additional knowledge that is encoded into the documents' metadata and inter-document relationships, like author names belonging to a couple of relevant documents. In our approach we explicitly make use of the user's intention to get an answer for his "who" interest. We use co-occurrence analysis methods (Jaccard index) to compute a semantic distance between terms from title/abstract and the co-occurring authors and thesaurus terms in the whole document set. Using the Jaccard similarity measure we are able to compute the most similar authors and thesauri terms to a topic expressed through 1..n terms. In our social sciences setting, doing a query on *retirement* and *health* results in names like *Richard Hauser* or *Gerhard Bäcker*. Both of them are social scientists who did work related to social welfare and retirement topics. Additionally we get thesaurus terms like *social politics* or *elderly people*. We compute the top n=4 associated author names and thesaurus terms and expand the query with these recommendations. In addition we extend the search on different fields like title and abstract (TI, AB), controlled term (CT) and authors (AU).

The following example visualizes our approach (to keep the example readable we just add two names and terms respectively). An initial query on *Retirement and health issues*, would be transferred from a baseline query

```
TI/AB = (retirement OR health)
```

to the extended query where italic terms are expanded thesaurus term and the last line originated from the list of co-occurring author names:

```
TI/AB = (retirement OR health OR "social politics" OR "elderly people")
OR CT = (retirement OR health OR "social politics" OR "elderly people")
OR AU = ("Richard Hauser" OR "Gerhard Bäcker")
```

## 4      Evaluation

In our evaluation setting we established four different queries, which then form TF*IDF ranked result lists using the Solr search engine. The simple result set B is formed by the unexpanded query generated from the topic description without stop words. This value is our baseline to which all other queries are compared. The second and third queries are the baseline combined with a thesaurus term expansion (B+TE) and author names respectively (B+AE). The fourth query is a combination of all three (B+TE+AE). In our experimental setup we used the GIRT4 corpus and the CLEF topics 76-125 to evaluate the performance of the proposed methods. We calculated MAP, rPrecision, p@10, p@20 and p@100 using the trec_eval toolkit.

**Table 2.** Retrieval results of four different systems: baseline system B, baseline with term expansion (B+TE), baseline with author expansion (B+AE) and a combined system B+TA+TE. Statistical significant changes compared to B (measured using t-test with α = .05) are marked with *. Best values are marked with bold font.

| run | MAP | rPrecision | p@10 | p@20 | p@100 |
| --- | --- | --- | --- | --- | --- |
| B | 0,139 | 0,182 | 0,442 | 0,353 | 0,172 |
| B+TE | 0,153 | 0,229 * | 0,430 | 0,399 | 0,217 * |
| B+AE | 0,122 * | 0,184 | 0,400 * | 0,350 | 0,175 |
| B+TE+AE | **0,170** * | **0,239** * | **0,478** | **0,427** * | **0,218** * |

The results of our evaluation are listed in table 2. The traditional approach B+TE, which adds controlled terms from a thesaurus to the query leads to a slight but not statistical significant increase in MAP (+9%) and a significant increase in rPrecision and p@100. B+AE performed significantly worse compared to B and B+TE, while rPrecision, p@20 and p@100 are comparable to the baseline. MAP (-12%) and p@10 are clearly below the values from B and B+TE. The combined method B+TE+AE performs best in all reported values. Generally speaking the B+AE approach in most cases was significantly worse than any other approach but combined with B+TE lead to significant gain in MAP, rPrecision, p@20 and p@100. When choosing B+TE as the common baseline the increase in B+TE+AE's MAP (+11%) is still significant.

## 5    Conclusion and Future Work

Looking back on the applications of the principle of polyrepresentation we see its use in the form of data fusion or ranking influential factors [10] rather than in the domain of query expansion and term suggestion. In this study we sketched out a term suggestion module that not only recommends controlled vocabulary terms but also the names of topical-related authors. We have shown that the explicit combined use of functional and cognitive polyrepresentational document features can increase the retrieval quality of a system. This is in line with traditional findings in the domain of query expansion but we could show that by using additional query entities that real users of a DL system were looking for, could bring a measurable benefit. This effect is only visible when the query is additionally expanded with thesaurus terms. A clear indicator for the strength of our polyrepresentational approach is the fact, that by only adding the author names we can observe a query drift which results in significantly worse results.

With our co-occurrence-based approach we can suggest other entities encoded in our digital library system. In this paper we focused on co-occurring and therefore potentially topic-related authors but in general other entities like journals, publisher, affiliation or other scientific community-related entities could be used to reformulate and enhance the original query. The open research questions would be: Are related journals, affiliations or any other DL's metadata capable of adding a positive retrieval effect or maybe a different view onto the data set? How can search effectiveness benefit from these co-occurrences? Does the polyrepresentational principle hold for these entities? While in this work we could only show that this principle holds for thesaurus terms (for a more detailed study on thesaurus terms see work by Lüke et al. in these proceedings) and authors names a more general evaluation and study on this has to remain as future work.

**Acknowledgements.** This work was partly funded by DFG, grant no. SU 647/5-2.